\documentclass[aps,pre,twocolumn,groupedaddress,showpacs,floatfix]{revtex4}
\usepackage{graphicx}
\usepackage{pst-all}
\usepackage{color}
\usepackage{amsmath}
\newcommand{\comment}[1]{}

\begin{document}
\title{On the properties of cycles of simple Boolean networks}
\author{Viktor Kaufman and Barbara Drossel}
\affiliation{Institut f\"ur Festk\"orperphysik,  TU Darmstadt,
Hochschulstra\ss e 6, 64289 Darmstadt, Germany }
\date{\today}
\begin{abstract}
We study two types of simple Boolean networks, namely two loops with a
cross-link and one loop with an additional internal link. Such
networks occur as relevant components of critical $K=2$ Kauffman
networks. We determine mostly analytically the numbers and lengths of
cycles of these networks and find many of the features that have been
observed in Kauffman networks. In particular,
the mean number and length of cycles can diverge faster than any power
law. 
\end{abstract}
\maketitle

\section{Introduction}
\label{sec:intro}

The study of random Boolean networks is of great interest since these
networks are one of the simplest models of genetic regulatory
networks.  Although they were introcuded already 40 years ago by
Kauffman\cite{Ka1}, they are still poorly understood. Due to
increasing computational power, it was recently discovered that old
assumptions about the properties of the cycles of these networks
have been wrong, see \cite{BaPa1,SK,BS}. To better understand Boolean networks is an important
requirement before being able to study successfully more realistic,
but also more complicated models.

Random Boolean networks are directed graphs consisting of $N$ binary
nodes, each having inputs from $K$ randomly chosen other nodes. To
each node, a Boolean function is assigned that gives the updating rule
of the node as function of input values. The network is updated
synchronously, and starting from an initial state, the network
eventually reaches a periodic trajectory (a cycle).  The situation $K=2$
is particularly interesting since it is the critical point between the
ordered regime (where only a finite number of nodes are not frozen for
$N \to \infty$) and chaos (where a small perturbation spreads through
the entire network). For this reason, it was believed for a long time
that the number and mean length of cycles of critical networks
increases as a power law with network size $N$. However, 
recent computer simulations \cite{SK} as well as analytical calculations
\cite{sam03}
indicate that the number of cycles of critical Boolean
networks increases faster than any power law with $N$.
So far, none of these studies provides direct intuitive insights in
how this feature emerges from the network structure.  Additionally, 
there is yet little agreement on the behavior of the mean length
of cycles.

This work aims at understanding better how such vast numbers and sizes
of cycles can emerge. For this purpose, we refer to the concept of
\emph{relevant nodes} introduced in \cite{BaPa2}. These are those
nodes of the network that can influence themselves via a loop of
connections. Their state undergoes therefore a nonconstant sequence of
values at least on some cycles. The network that remains after
removing the irrelevant nodes consists only of loops and links between
and inside loops. Most nodes are frozen \cite{Fly}, and recent work
\cite{SK} suggests that the number of relevant nodes increases as
$N^{1/3}$ with the number of nodes. The reduced network that contains
only the relevant nodes determines the number and lengths of cycles in
the full network. It has only slightly more than one input per
node. However, little is yet known about the number of cycles even on
the simplest possible relevant networks, apart from simple loops. We
therefore focus in this article on the cycles of two of the simplest
relevant networks (or building blocks of relevant networks) that can
be constructed, namely two cross-linked loops and one loop with an
additional link within the loop. We shall see that the mean number of
cycles on these simplest relevant networks increases faster than any
power law with the number of nodes of these networks, and for some of
these networks also the mean cycle length increases faster than any
power law.  Since it can be expected that these simple networks occur
within the relevant network of critical Boolean networks, we now
understand better properties of cycles in critical networks.

The outline of this paper is as follows: In the next section, we
briefly review the properties of cycles on simple loops. In section
\ref{sec:twoloops}, we study the cycles of two cross-linked loops.  In
section \ref{sec:extralinkloop}, we focus on loops with one additional
link, and in the final section, we discuss our results.

\section{Simple loops}
\label{sec:loops}

Trivial loops consisting of $N$ nodes are the simplest networks. Each
node has one input, just as in a $K=1$ network, and the nodes are
connected to form a loop. Since we are only interested in systems
consisting of relevant nodes, we consider the case where out of the 4
possible Boolean functions only the two nontrivial ones occur. These
are ``truth'', which simply copies the value of the input at the
update, and the Boolean negation.

A loop with $n$ negations can be mapped bijectively onto a loop with
$n-2$ negations by replacing the two negations with truth and by
inverting the state of all nodes between these two links. For this
reason, we need to consider only loops with zero negations and loops
with one negation. We refer to these two situations as the ``even''
and ``odd'' case respectively. The dynamics on these loops has the
following obvious properties, \hbox{see also \cite{ACK}, \cite{kja88_2}:}
\begin{enumerate}
\item After $N$ updates, a loop with an even number of negations
  returns to the same state. A loop with an odd number of negations
  returns to the same state after $2N$ updates.
\item Consequently, each state is on a cycle, and the mean
  cycle length, multiplied by the number of cycles, is $2^N$. 
\item No cycle can be longer than $N$ (even) or $2N$ (odd). Loops
  with zero negations have 2 fixed points (all 1 or all 0), and loops
  with one negation have a cycle of length 2 (alternating 0 and
  1).
\item  If $N$ is a prime number, the number of cycles is given by 
\begin{equation}
C_N = \left\{
\begin{array}{ll}
\;2+\frac{2^N-2}{N}& \text{\ even case}\\
\;1+\frac{2^N-2}{2N}& \text{\ odd case}
\end{array}
\right.
\label{eq:C_N_triv}
\end{equation}
This result does not apply to an odd two-node system $N=2$. In this
case, there is one cycle that comprises all 4 states. 
\item If $N$ is not a prime
  number, any divisor of $N$ ($2N$) is also a cycle length. There exist more shorter cycles,
  and therefore the number of cycles is larger than the above
  expression.
\end{enumerate}
To summarize, simple loops have a mean cycle length of the order
of $N$, and an average number of cycles that increases as $2^N/N$,
which is faster than any power law in $N$. 

\section{Two loops with cross-link}
\label{sec:twoloops}

\begin{figure}
\vskip 2cm
\begin{center} \setlength{\unitlength}{1pt} 
\begin{picture}(0,0)
\psset{unit=1pt,fillcolor=black}
\pscircle(-60,0){30}
\pscircle(60,0){30}
\psline{->}(-26,0)(26,0)
\pscircle(30,0){3}
\rput[lt]( 36,5){$\Sigma$}
\pscircle(34,-15){3}
\psline{->}(31,-6)(30,-3)
\rput[lt](40,-10){$G_2$}
\pscircle(-30,0){3}
\psline{->}(-33,-13)(-34,-15)
\rput[lb](-26,4){$G_1$}
\rput[cc](-96,20){$N_1$}
\rput[cc]( 96,20){$N_2$}
\end{picture}
\end{center}
\vspace*{+4.0ex}
\caption{Two loops with a cross-link.}
\label{fig:twoloops}
\end{figure}
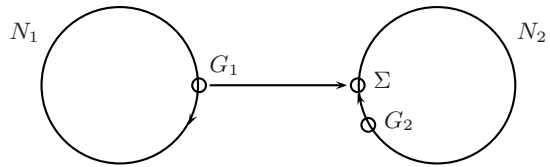

We next consider two loops of size $N_1$ and $N_2$ with a cross-link
(see Fig.~\ref{fig:twoloops}).  We denote with $\Sigma$ the node with
two inputs, and with $G_1$ and $G_2$ the two nodes it receives its
input from. Again, we consider only the case where all links are
relevant. Without loss of generality, the first loop has only truth
functions or one negation. The second loop has truth functions at all
nodes apart from $\Sigma$, and one of the following three Boolean
functions at $\Sigma$: $f_{11}$, which is $0$ if and only if $G_1=0$
and $G_2=1$; $f_{14}$, which is $0$ if and only if both its inputs are
$0$; and finally the function $f_9$, which takes the value $G_2$ if
$G_1=1$ and the inverted value of $G_2$ if $G_1=0$. The first two
functions are canalyzing functions. This means that there exists at
least one input configuration for which inverting one input does not
change the output.  The third function is reversible, since to each
state the network has a unique predecessor. Each state of the system
is therefore on a cycle, and the mean cycle length, multiplied by the
number of cycles, is $2^{N_1+N_2}$. The other four canalyzing
functions and the second reversible function need not be considered,
since networks with these functions can be mapped on networks with the
given three functions by inverting the states of all nodes in the
first loop, or by inverting the states of all nodes.

\subsection{Case 1: $N_1$ and $N_2$ are prime numbers}
\label{subsec:prime}

In the following, we focus on the case that $N_1$ and $N_2$ are prime
numbers (with $N_1 \neq N_2$). The first loop provides a periodic
input to $\Sigma$ of the period $p_1=N_1$ or $2N_1$ or 1 or 2. Loop 2
behaves like a single loop, where the Boolean function at $\Sigma$
changes between truth and negation ($f_9$) according to a pattern of
period $p_1$, or between negation and 1 ($f_{11}$), or between truth
and 1 ($f_{14}$). Loop 2 returns to the same state no later than after
$2p_1N_2$ updates. The largest cycle has therefore the length
$4N_1N_2$ (except for $N_2=2$, where the largest cycle has the length
$4N_1$).

If the Boolean function at $\Sigma$ is canalyzing, most results can be
derived from the observation that for $p_1>1$ the first input to
$\Sigma$ is 1 every $2N_1$ time steps, and possibly more often.  Let
us first consider the function $f_{14}$.  A 1 at $G_2$ will lead again
to a 1 at $G_2$ after $nN_2$ updates, for any integer $n$. A 0 at
$G_2$, combined with a 0 at $G_1$, will lead to a 0 at $G_2$ after
$N_2$ updates.  However, at the latest after $n=2N_1$ update cycles of length
$N_2$, this 0 will become a 1. Therefore, loop 2 will be frozen to all
1 after this time. The cycles of the network have length $p_1$ if
$p_1>1$.  If $p_1=1$, we obtain cycles of length $N_2$ and 1. We
conclude that if there is a function $f_{14}$ at $\Sigma$ and no
negation in loop 1, the system has $(2^{N_2}-2)/N_2$ cycles of length
$N_2$, three cycles of length 1 and $(2^{N_1}-2)/N_1$ cycles of length
$N_1$. If there is a function $f_{14}$ at $\Sigma$ and one negation in
loop 1, we have one cycle of length 2 and $(2^{N_1}-1)/2N_1$ cycles of
length  $2N_1$.

Next, let us consider the function $f_{11}$ at $\Sigma$. If $p_1=1$
and loop 1 is in state 1, the entire system is frozen in state 1. If
loop 1 is in state 0, loop 2 is like an independent loop with one
negation. If $p_1=2$, the entire system has period 2. If $p_1=N_1$ or
$2N_1$, the first loop enslaves the second loop, completely
determining its state and wiping out every memory of its initial
state. Consequently, the cycle length is $N_1$ ($2N_1$) for an
even (odd) first loop. We conclude that if there is a function
$f_{11}$ at $\Sigma$ and no negation in loop 1, there is one cycle
of length 1, one cycle of length 2, $(2^{N_2}-2)/2N_2$ cycles of length $2N_2$, and
$(2^{N_1}-2)/N_1$ cycles of length $N_1$. If loop 1 has one
negation, there is one cycle of length 2 and $(2^{N_1}-2)/2N_1$
cycles of length $2N_1$.

To summarize so far, the  number of
cycles for a system with $N_1$ and $N_2$ being prime numbers and
with a canalyzing function at $\Sigma$ is
\begin{eqnarray}
\label{eq:C_n_prime_canal}
C^{f_{14}}_{N_1, N_2} &=& 
\left\{
\begin{array}{ll}
\;3+\frac{2^{N_1}-2}{N_1} + \frac{2^{N_2}-2}{N_2}& \text{\ \!\!\!\!even loop 1}\\
\;1 + \frac{2^{N_1}-2}{2N_1}& \text{\ \!\!\!\!odd loop 1}
\end{array}
\right.\nonumber \\
C^{f_{11}}_{N_1, N_2} &=& 
\left\{
\begin{array}{ll}
\;2+ \frac{2^{N_1}-2}{N_1} + \frac{2^{N_2}-2}{2N_2}& \text{\ \!\!\!\!even loop 1}\\
\;1 + \frac{2^{N_1}-2}{2N_1} & \text{\ \!\!\!\!odd loop 1}
\end{array}
\right.
\end{eqnarray}
(These equations are modified if one of the loop sizes is $N_i=2$. The
terms with $2N_i$ in the denominator then have to be dropped.)
For large $N_1$ and $N_2$ the mean number of
cycles grows
as  $2^{N_{\text{max}}}/N_{\text{max}}$ with
$N_{\text{max}}$ being the larger of the two loop sizes, and the mean cycle
length increases linearly with $N_{\text{max}}$.

Finally, let us consider the function $f_{9}$ at $\Sigma$. If an even
loop 1 is frozen in state 1 (0), loop 2 behaves like an even (odd)
independent loop.  We get 2 fixed points (one cycle of length 2) for
the entire network and $(2^{N_2}-2)/N_2$ cycles of length $N_2$
($(2^{N_2}-2)/2N_2$ cycles of length $2N_2$). If an odd loop 1 is on
the cycle of length 2, the two loops have one cycle of length 4 and
$(2^{N_2}-2)/2N_2$ cycles of length $4N_2$

If loop 1 has period $p_1=N_1>1$ with an even number of 0s, the state
of $G_2$ will be the same every $N_1N_2$ time steps.  For a given
cycle with period $N_1$ on loop 1, two cycles with period $N_1$ of the
entire system can be constructed in this case as follows.  Begin by
fixing the initial value of one node in loop 2. After one time step,
the next node on loop 2 (in clockwise direction) will be the one that
is fixed, etc. Update the system for $N_1$ time steps and observe the
value that will be fixed then, and choose this to be the initial state
of that node. After iterating this procedure $N_2$ times, one has
fixed the initial state of all $N_2$ nodes, and one returns to the
initial node. Due to the even number of zeros on loop 1, the initial
node will then have again its initial value. We have thus created an
initial state that lies on a cycle of length $N_1$. A second cycle of
length $N_1$ is created by starting with the second possible initial
value.  All other cycles have the period $N_1N_2$.

If loop 1 has period $p_1>1$ with an odd number of 0s, which is always
the case for an odd loop 1, the state of $G_2$ will be the same every
$2p_1N_2$ time steps.  For a given cycle with period $p_1$ on loop 1,
a cycle with period $2p_1$ of the entire system can be constructed as
above, since two subsequent periods of loop 1 have an even number of
0s.  The other cycles have length $2N_2p_1$.

Our considerations lead to the following numbers and lengths of cycles
in systems with a reversible function at $\Sigma$:
\begin{center}
\begin{tabular}{c|c|c|c|c|c|c}
length& 1& 2& $N_2$& $2N_2$& $N_1$&$2N_1$\\
\hline
number & 2&1&$\frac{2^{N_2}-2}{N_2}$&$\frac{2^{N_2}-2}{2N_2}$&$\frac{2^{N_1}-2}
{N_1}$&$\frac{2^{N_1}-2}{2N_1}$
\end{tabular}
\begin{tabular}{c|c|c}
length& $N_1N_2$& $2N_1N_2$\\
\hline
number &$\frac{(2^{N_1}-2)(2^{N_2}-2)}{2N_1N_2}$&$\frac{(2^{N_1}-2)(2^{N_2}-2)}{4N_1N_2}$
\end{tabular}
\end{center}
for an even loop 1, and
\begin{center}
\begin{tabular}{c|c|c|c|c}
length & 4&$4N_2$& $4N_1$ & $4N_1N_2$ \\
\hline
number&1&$\frac{2^{N_2}-2}{2N_2}$&$\frac{2^{N_1}-2}{2N_1}$&$\frac{(2^{N_1}-2)(2^{N_2}-2)}{4N_1N_2}$
\end{tabular}
\end{center}
for an odd loop 1.  (Again, the results are modified if a loop has
size 2. For $N_1=2$ and an even loop 1, there is no cycle of length
$N_1$ or $N_1N_2$, and the cycles of length $2N_1$ and $2N_1N_2$ occur
twice as often. For an odd loop 1, the first two columns vanish, and
the other two cycle numbers are doubled. For $N_2=2$ and an even
loop 1, column 3,5,7 vanish, the cycle numbers in column 4,6,8 are
doubled. For an odd loop 1, column 1 and 3 vanish, and the other
cycle numbers are doubled.)

The mean number of cycles diverges as 
\begin{equation}
\label{eq:C_n_prime_revers}
C^{f_9}_{N_1, N_2} \simeq 
\left\{
\begin{array}{ll}
\;\frac{3\cdot 2^{N_1+N_2}}{4N_1N_2}& \text{\ even loop 1}\\
\; \frac{2^{N_1+N_2}}{4N_1N_2}& \text{\ odd loop 1}
\end{array}
\right.
\end{equation}
and the mean cycle length increases as $N_1N_2$.  Apart from the
prefactor, this result is the same as for two uncoupled loops.

\subsection{Case 2: $N_1=N_2\equiv N$ } 

We call this case ``resonant'', because here one has substantially
more cycles for canalyzing $f$s in comparison to the case $N_1\neq
N_2$ with $N_1$, $N_2$ of the same order of magnitude. Since each node
value of loop 2 can be changed at $\Sigma$ by exactly one node value
of loop 1, the system can be decomposed into $N$ independent systems
of 2 nodes, where the first node receives input from itself (negation
for an odd loop 1, otherwise truth function), and the second node
receives input from both nodes.  These $N$ systems are updated one
after another. If the first loop is even and the Boolean function at
$\Sigma$ is $f_{14}$, the 2-node system has three cycles of length
1. The complete system has therefore 3 cycles of length 1 and 
$\frac{3^{N}-3}{N}-\delta_{N,2}$ cycles of length $N$.
 
If the first loop is odd and the Boolean function at $\Sigma$ is
$f_{14}$, the 2-node system has one cycle of length 2.  The complete
system has therefore one cycle of length 2 and $\frac{2^{N}-2}{2N}$
cycles of length $2N$. The first loop enslaves the second loop. (For
$N=2$, there is only one cycle of length 4.)

If the first loop is even and the Boolean function at $\Sigma$ is
$f_{11}$, the 2-node system has one cycle of length 1 and 1 cycle of
length 2.  The complete system has therefore one cycle of length 1,
one cycle of length 2, and $\frac{3^{N}-3}{2N}$ cycles of length $2N$.
(For $N=2$, there are only two cycles of length 4.)

If the first loop is odd and the Boolean function at $\Sigma$ is
$f_{11}$, the first loop enslaves the second loop.  The complete
system has therefore one cycle of length 2 and $\frac{2^{N}-2}{2N}$
cycles of length $2N$. (For $N=2$, there is only one cycle of length
4.)

If the first loop is even and the Boolean function at $\Sigma$ is
$f_{9}$, the 2-node system has two cycles of length 1 and one cycle of
length 2.  The complete system has therefore two cycles of length 1,
one cycle of length 2, $\frac{2^{N}-2}{N}$ cycles of length $N$ (none
for $N=2$), and $\frac{4^{N}-2^N-2}{2N}$ (3 for $N=2$) cycles of
length $2N$.

If the first loop is odd and the Boolean function at $\Sigma$ is
$f_{9}$, the 2-node system has one cycle of period 4.  The complete
system has therefore one cycle of period 4 and $\frac{4^{N}-4}{4N}$
cycles of period $4N$. (For $N=2$, there are only two cycles of length
8.)

For large $N$, the number of cycles diverges as
\begin{eqnarray}
\label{eq:C_nn_prime}
C^{f_9}_{N, N} \simeq \frac{4^{N}}{2N} \text{ or }\frac{4^{N}}{4N}\nonumber\\
C^{f_{14}}_{N, N} \simeq \frac{3^N}{N} \text{ or }\frac{2^N}{2N}\\
C^{f_{11}}_{N, N} \simeq \frac{3^N}{2N} \text{ or }\frac{2^N}{2N}\nonumber
\end{eqnarray}
for an even or odd first loop,
and the mean cycle length increases linearly in $N$. 
Our computer simulations are in agreement with the analytical results.

\subsection{Case 3: General $N_1$ and $N_2$ }
\label{subsec:general}

If $N_1$ and/or $N_2$ are not prime numbers, there are more
cycles. First, let us consider the case that $N_1$ and $N_2$ have no
common divisor and that loop 1 is even if $N_2$ is even. The above
listed cycle lengths 1, 2, $N_1$, $N_2$, $2N_1$, $2N_2$, $4N_1$,
$4N_2$, $N_1N_2$, $2N_1N_2$, $4N_1N_2$ still occur, but there exist
additional cycle lengths, which are obtained by replacing $N_1$ and/or
$N_2$ with one of its divisors. The numbers of cycles with lengths
from the list will decrease accordingly.  

\begin{figure}
\begin{center}
\includegraphics[width=0.62\columnwidth]{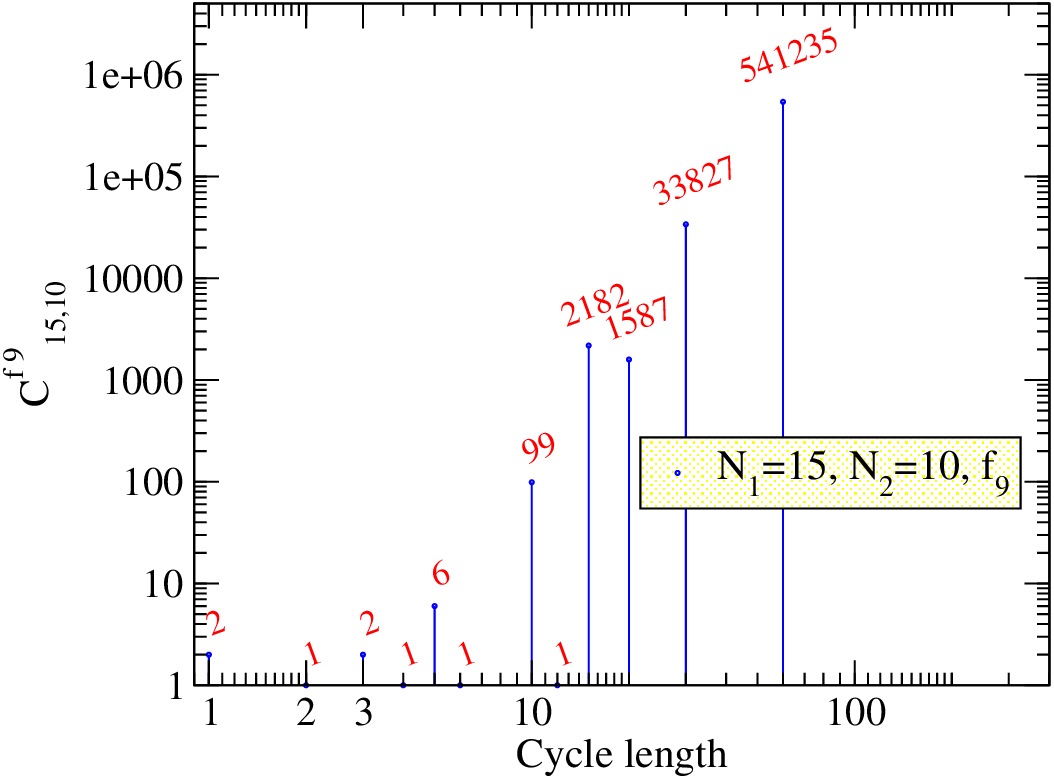}\\
\includegraphics[width=0.62\columnwidth]{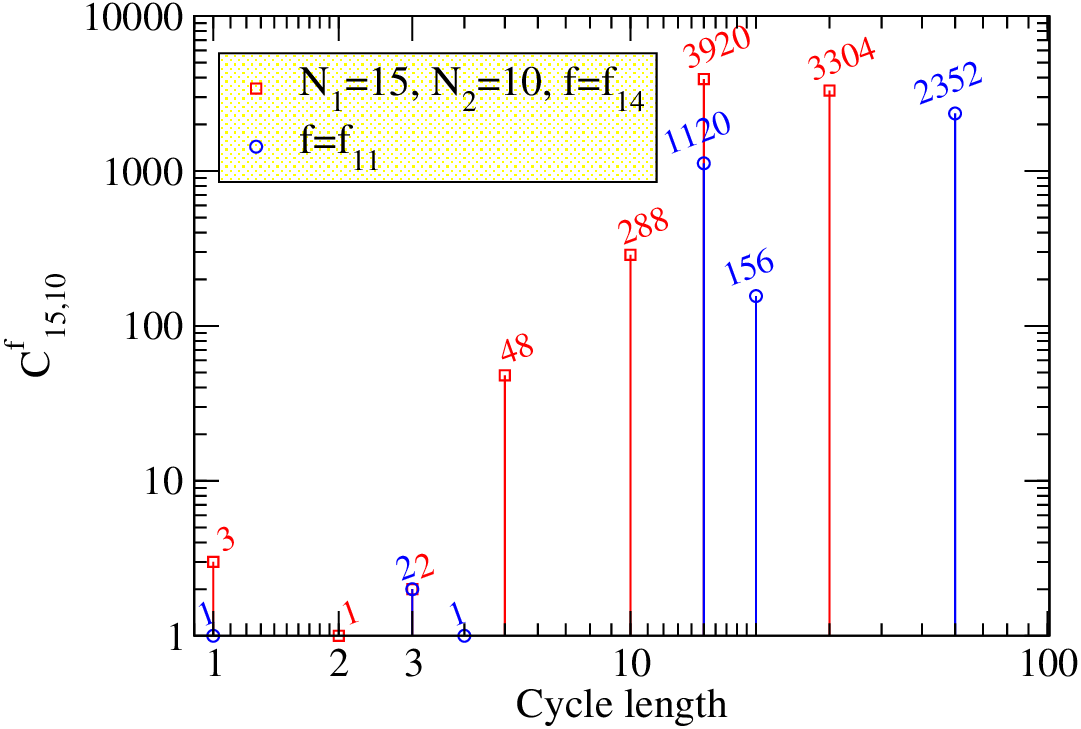}\\
\includegraphics[width=0.62\columnwidth]{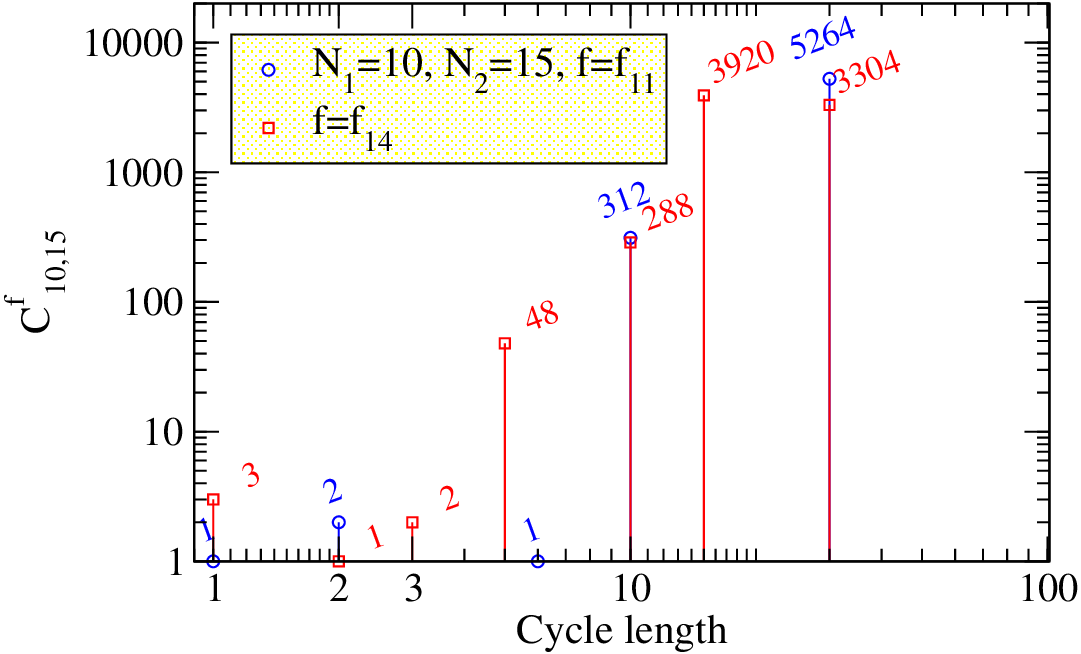}
\end{center}
\caption{Three examples of numerical results for the number of cycles as function of their length for two loops with a cross-link, with an even first loop}
\label{fig:f_10_15}
\end{figure}

In the remainder of this section we consider the more interesting case
that the cycle length of loop 1, $P_1$, and $N_2$ have a greatest
common divisor $g\equiv g(P_1)>1$. This is always the case if $N_1$
and $N_2$ have a common divisor. The special case $N_1=N_2$ was treated in
the previous subsection.

The least common multiple of $P_1$ and $N_2$ is $P_1N_2/g$ and, for a
given $P_1$, the largest possible cycle length is $2P_1N_2/g$ and the
smallest possible cycle length is $P_1$. The values of one period of
loop 1 and the nodes of the second loop split into $g$ independent
subsystems with $P_1/g$ values in each periodic sequence from loop 1
at $G_1$, and $N_2/g$ nodes from the second loop.  One subsystem is
updated at a time and takes place of the next one in the sequence.
For a handy picture of the subsystems one can imagine the sequence of
period $P_1/g$ as being produced by an even loop with $P_1/g$
nodes. In the case of an odd loop 1 and an even $P_1/g$, the second half
of the period of such a new loop 1 in a subsystem is the inversion of
the first half. In the case of an odd loop 1 and an odd $P_1/g$ the
subsystems come in pairs; to each subsystem with an odd number of 0s
in the periodic sequence from loop 1 there exists a subsystem with an
even number of 0s. The 0s and 1s are interchanged. We call these
subsystems complementary.

The numbers and lengths of cycles of a subsystem can be calculated
according to the rules outlined in the previous subsections.  Let us
now point out some rules that help determining the possible cycle
lengths of the entire system if the cycles in the subsystems are
given, their lengths be denoted by $p_1,p_2,\dots$. If each subsystem
is on a different cycle, the cycle length of the entire system is
$T={\rm LCM}(p_1g,p_2g,\dots)$. LCM stands for the least common
multiple. Otherwise shorter cycles can exist. For example, if all
subsystems are on the same cycle, $p_1=p_2=\dots= p$, the phase shifts
between subsystems can be arranged in such a way that overall periods
shorter than $pg$ occur. These periods can be any divisor of $pg$ that
is a multiple of $p$, but not a multiple of $g$.

Now, let us turn to the number of cycles. We first consider the
reversible Boolean function $f_9$ at $\Sigma$. If $N_1$ and $N_2$ are
large and for an even first loop, it is sufficient to consider
$P_1=N_1$, so that each subsystem is approximately with probability
0.5 on a cycle of length $N_1/g\cdot N_2/g$ and with probability 0.5
on a cycle of length $N_1/g\cdot 2N_2/g$. The subsystems are almost
certainly on different cycles. The probability that the overall cycle
length is $N_1N_2/g$ is $0.5^g$, and the probability that the overall
cycle length is $2N_1N_2/g$ is $(1-0.5^g)$. We can neglect the cycles
of length $N_1N_2/g$, since their number is $0.5^g/(1-0.5^g)$
times smaller than that of the cycles of length $2N_1N_2/g$. The next
neglected contributions to the number of cycles would be from
cycles of lengths $2N_2$,$N_2$,$2N_1$,$N_1$.  

If the first loop is odd, we can restrict ourselves to looking at
$P_1=2N_1$. Each subsystem or each pair of complementary subsystems is
with probability near to 1 on a cycle of length $2\cdot 2N_1/g\cdot
N_2/g$, and the overall cycle length is $4N_1N_2/g$. In our estimation
for the number of cycles the most significant contributions we neglect
come from cycles of lengths $2P_1$ with $P_1=2N_1$ and $4N_2/g$ with
$P_1=2$.  Equation (\ref{eq:C_n_prime_revers}) for the mean number of
cycles for large $N_1$ and $N_2$ becomes now
\begin{equation}
C^{f_9}_{N_1, N_2} \simeq 
\left\{
\begin{array}{ll}
\;\frac g 2 \frac{2^{N_1+N_2}}{N_1N_2}& \text{\ even loop 1}\\
\; \frac g 4 \frac{2^{N_1+N_2}}{N_1N_2}& \text{\ odd loop 1}
\end{array}
\right.
\label{eq:C_n_commens_revers}
\end{equation}

For canalyzing Boolean functions, there is now a big difference
between the case of an even loop 1 and an odd loop 1.  If loop 1 is
odd for odd $N_2$ it always enslaves the second loop, and the value of
$N_2$ does not matter. We obtain no new results beyond what has been
written in the previous subsections. The majority of cycles have
length $2N_1$. Their numer is of the order of $2^{N_1}/2N_1$.  We
obtain this and the following results systematically by combining the
results for individual subsytems.  For instance, for even $N_2$ and
$P_1=2$ one of the two subsystems is all 1 and the other one is all
0. For the function $f_{14}$ at $\Sigma$ we then get of the order of
$2^{N_2/2}/(N_2/2)$ cycles of length $N_2$. For $f_{11}$ we get of the
order of $2^{N_2/2}/N_2$ cycles of length $2N_2$.

For an even loop 1 the change in cycle size and number
is dramatic compared to the case where $N_1$ and $N_2$ are prime
numbers. In particular, cycles of lengths $N_1N_2/g$ and
$2N_1N_2/g$ appear now, since some subsystems may have the period
$N_1/g$ and some subsystems the period $N_2/g$. Let us first consider
the function $f_{14}$. For large $N_1$ and $N_2$,
each subsystem is almost certainly in one out of approximately
$2^{N_1/g}$ states belonging to cycles of length $N_1/g$ or in one out
of $2^{N_2/g}$ states belonging to cycles of length $N_2/g$.  The 
number of cycles for large $N_1$ and $N_2$ is therefore
\begin{equation}
C^{f_{14}}_{N_1, N_2} \simeq 
\frac g {N_1N_2} \left(2^{N_1/g} + 2^{N_2/g}\right)^g\,.
\label{eq:C_n_commens_f14}
\end{equation}%
A more detailed treatment leads to the following expression for this quantity
\begin{gather}
\label{eq:C_n_commens_f14_full}
C^{f_{14}}_{N_1, N_2} \simeq 
\frac {g  \left(2^{N_1/g} + 2^{N_2/g}-1\right)^g}{N_1N_2}+\\\nonumber%
+\frac{\left(2^{N_1/g}+1\right)^g}{N_1}+\frac{\left(2^{N_2/g}+1\right)^g}{N_2}\,,
\end{gather}%
where the dominant cycles of the lengths $N_1N_2/g$, $N_1$ and $N_2$ have been taken into accout.

Finally, let us consider the Boolean function $f_{11}$.  If $N_1/g$ is
even, the longest cycle length is $N_1N_2/g$, otherwise it is
$2N_1N_2/g$. We have therefore
\begin{equation}
C^{f_{11}}_{N_1, N_2} \simeq 
\left\{
\begin{array}{l}
\;\frac {g \left(2^{N_1/g}+2^{N_2/g}-1\right)^g}{N_1N_2}+\frac{\left(2^{N_1/g}+1\right)^g}{N_1}+\\
\quad+\frac{\left(2^{N_2/g}+1\right)^g}{2N_2}\qquad\quad \text{\ even $N_1/g$}\\
\\
\;\frac {g \left(2^{N_1/g}+2^{N_2/g}-1\right)^g}{2N_1N_2}+\frac{\left(2^{N_1/g}+1\right)^g}{2N_1}+\\
\quad+\frac{\left(2^{N_2/g}+1\right)^g}{2N_2}\qquad\quad \text{\ odd $N_1/g$}
\end{array}
\right.
\label{eq:C_n_commens_f11}
\end{equation}

As an illustration of the findings of this subsection, we show in
figure \ref{fig:f_10_15} the results of three numerical evaluations of
the cycles of a two-loop system with $g=5$. Compared to two
independent loops, for which the largest cycle length is $N_1N_2/g$,
the largest cycle can now have up to four times this length. When the
Boolean function at $\Sigma$ is canalyzing, the cycles are comparatively
shorter and there are more of them. In any case there exist characteristic
dominant cycle lengths.
The total number of cycles increases faster than any
power law with $N_1$ and $N_2$, but the mean cycle length increases
linearly in $N_1$ and $N_2$. 

\section{Loops with one additional link}
\label{sec:extralinkloop}
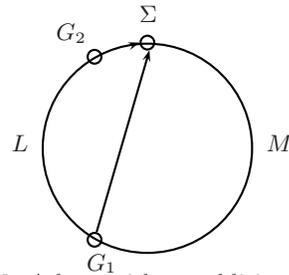
\begin{figure}
\vskip 2cm
\begin{center} \setlength{\unitlength}{1pt} 
\begin{picture}(0,0)
\psset{unit=1pt,fillcolor=black}
\pscircle(0,0){40}

\pscircle(-20,34.64){3}
\psline{->}(-6,39)(-3,40)
\pscircle(0,40){3}
\rput[lb]( -3,48){$\Sigma$}

\pscircle(-20,-34.64){3}
\psline{->}(-19.22,-31.74)(0.78,37.1)

\rput[rt](-45,5){$L$}
\rput[lt](45,5){$M$}
\rput[rb](-23,40){$G_2$}
\rput[lt](-23,-40){$G_1$}

\end{picture}
\end{center}
\vspace*{+6.0ex}
\caption{A loop with an additional link}
\label{fig:extralinkloop}
\end{figure}

Now let us turn to a loop of size $N=L+M+2$ with one additional link,
 as shown in figure \ref{fig:extralinkloop}.  We denote with $\Sigma$
 the node with two inputs, and with $G_1$ and $G_2$ the two nodes it
 receives its input from. Again, we consider only the case where all
 links are relevant. Without loss of generality, we can assume that
 the Boolean functions at all nodes apart from $\Sigma$ are truth
 functions. At $\Sigma$, we shall consider the reversible function
 $f_9$, and the canalyzing functions $f_{14}$, $f_{11}$, $f_4$, and
 $f_{1}$. $f_4$ is 0 if the input from $G_1$ is 1, and otherwise it
 copies the values of the second input. $f_1$ yields 1 if and only if
 both inputs are 0. Systems with the other Boolean functions can be
 mapped on systems with these functions by inverting the states of all
 nodes. We count the nodes counterclockwise, assigning to $G_2$ the
 index $x=1$, to $G_1$ the index $x=L+1$, and to $\Sigma$ the index
 $x=N \equiv 0$. 

A system with $n<L$ nodes on the connection from $G_1$ to $\Sigma$ can
be mapped on the system shown in figure \ref{fig:extralinkloop} by
connecting node number $L+1-n$ directly to $\Sigma$.  In the
following, we will first consider the four canalysing functions, and
then the reversible function. We will use analytical calculations as
well as computer simulations. 

\subsection{Case 1: Boolean function $f_{14}$ at $\Sigma$}
\label{subsec:f14}

We first consider the simplest case, where an output $0$ is only
obtained if both inputs are $0$. Starting from a random initial
condition, the initial number of $0$s cannot increase. There are two
fixed points, all 0 and all 1.  Every $0$ needs another $0$ $L$ steps
back along the loop links in order to survive. Nontrivial cycles occur
only if the greatest common divisor of $N$ and $L$ is $g>1$. There are
then $g$ independent sets of nodes, which can be assigned a value $0$
or $1$. There are $2^g-2$ states on cycles of length $g$ or one of its
divisors.  The number of cycles, averaged over $L$ and over a small
interval of $N$ values increases at least as $2^{N/2}/N^2$ with $N$,
since for even $N$ and $L=N/2$ we have $g=N/2$.

\subsection{Case 2: Boolean function $f_4$ at $\Sigma$}
\label{subsec:f4}

\begin{figure}
\begin{center}
\includegraphics[width=0.85\columnwidth]{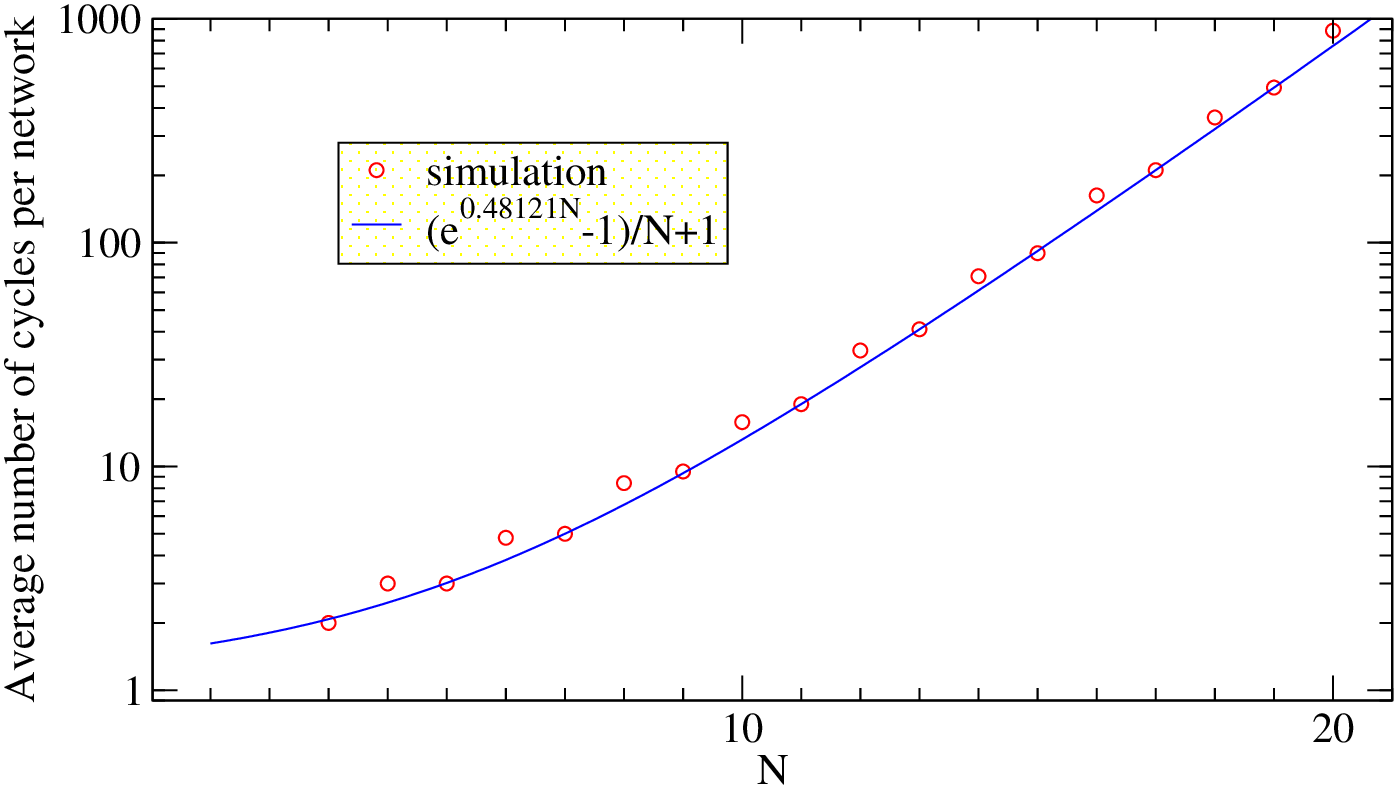}
\end{center}
\caption{Mean number of cycles per network for the canalyzing function
$f_4$ at $\Sigma$. }
\label{fig:f4_numattr}
\end{figure}

The next canalyzing function we consider yields a $0$ if the first
input is $1$, and copies the value of the second input otherwise.
Starting from a random initial condition, each node value $0$ comes
back to the starting location without change after one rotation (i.e.,
after $N$ time steps).  On a cycle, each 1 at $G_2$ must be followed
by a $0$ at $G_1$, $L$ nodes back, otherwise it would disappear as it
passes $\Sigma$.  Let us consider the sequence of states of $G_2$
every $L$ time steps on a cycle. If $g$ is the largest common divisor
of $L$ and $N$, there are $g$ independent sequences of length
$N/g$. For the number $\phi_N$ of different sequences of period $N$,
where each 1 is followed by a 0, one obtains the  recursive
equation
$$\phi_N=\phi_{N-1} + \phi_{N-2}\, ,$$ since a sequence of length $N$
can be obtained by adding a 0 after the first 1 of a sequence of
length $N-1$ (or at the end, if there is no 1) or by adding a 01 after
the first 1 of a sequence of length $N-2$ (or a 00 at the end, if
there is no 1).  The initial condition is $\phi_1=1$ and
$\phi_2=3$. For large $N$, we make the ansatz $\phi_N=a\cdot b^N$,
which leads to $b=(1+\sqrt 5)/2$. For $2\le N\lesssim 20$, we find
numerically $a=1$ using $b\simeq e^{0.48121}$. Consequently, if $N$
and $L$ have no common divisor, we expect the number of cycles to be
\begin{equation}
C^{f_{4}}_{N} \simeq \frac{e^{0.48121 N}-1}{N}+1\, .
\label{eq:C_n_extralink_f4}
\end{equation}
Otherwise, the number of cycles is somewhat larger. These results are confirmed
numerically, as shown in figure \ref{fig:f4_numattr}, where averaging over different $L$ has been performed.

\subsection{Case 3: Boolean function $f_1$ at $\Sigma$}
\label{subsec:f1}

Now we continue with a more complex case: the canalyzing function
$f_1$ yields $1$ if and only if both inputs are $0$. Consequently, if
one of the two inputs is 1, the output is 0. We will see that
there are again exponentially many cycles. 

\begin{figure}
\begin{center}
\vskip 1cm
\includegraphics*[width=0.85\columnwidth]{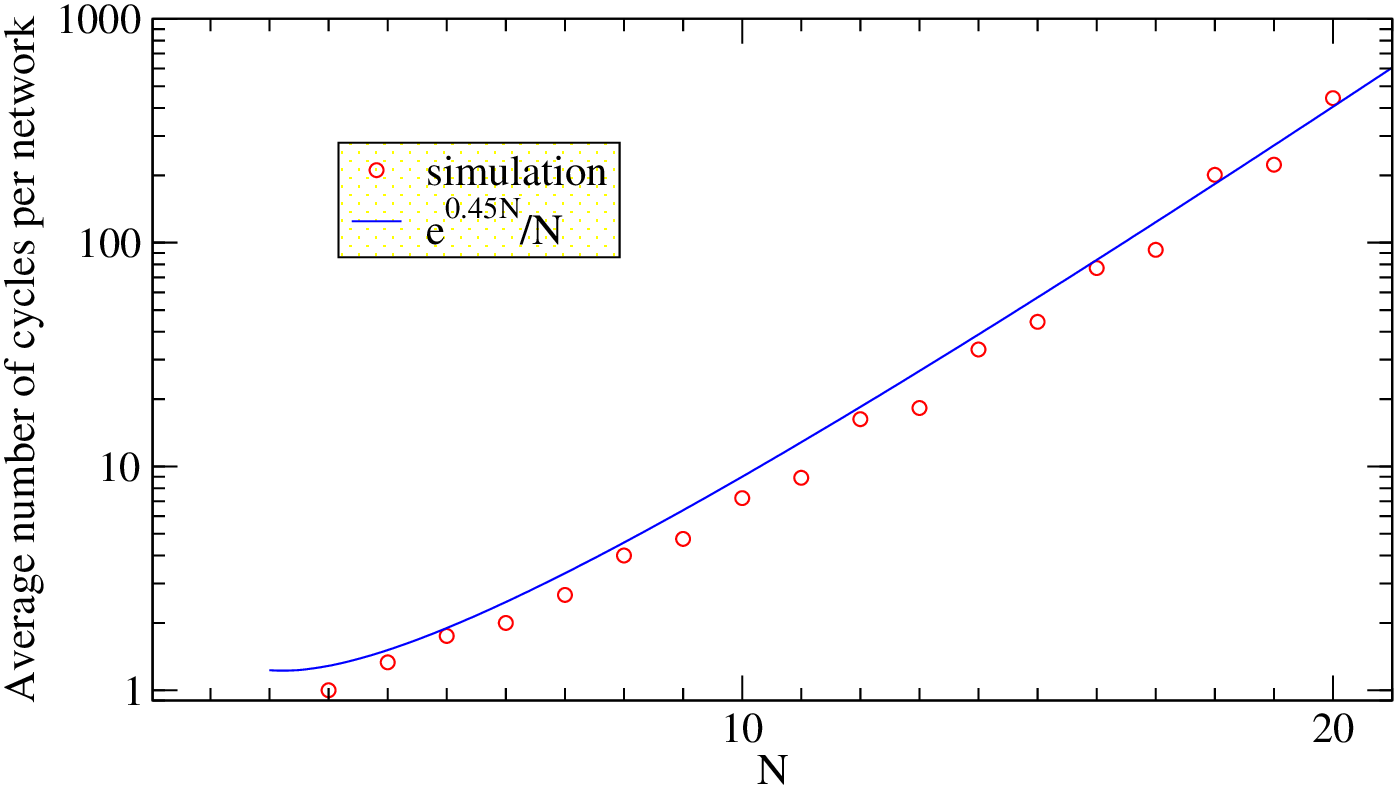}
\end{center}
\caption{Mean number of cycles per network for the canalyzing function
$f_8$ at $\Sigma$. }
\label{fig:f1_numattr}
\end{figure}

First, let us consider the fate of a node value $1$ on a cycle as we
iterate the network.  This 1 moves from site $x$ to site $x-1$ during
one time step. As it reaches the node $G_1$, it produces a $0$ at
$\Sigma$. When it reaches $G_2$, it produces another $0$
exactly $L$ sites behind the first one. These two zeros will produce a
value $1$ as soon as they reach the nodes $G_1$ and $G_2$
respectively. Thus, a $1$ comes back to its original place after
$2N-L$ steps.  In the same way, each pair of $0$s, $L$ steps apart
from each other, will come back to their original places after $2N-L$
steps. One can easily see that every 0 on a cycle must be a part of
such a pair: consider a 1 that has just been created at site
$\Sigma$. If after $L$ time steps there is a 0 at $\Sigma$, there must
be at the same time a 1 at $G_1$. After $L$ additional time steps,
there is consequently a 0 at $\Sigma$. We conclude that the period of
the cycles is $2N-L$ or one of its divisors. For $L=1$, the number of
cycles is equal to the number of sequences of length $N$, where 0s
always occur in pairs, with an appropriate boundary condition at $\Sigma$. 
The number of such sequences $\phi_N$ satisfies the recursion relation
\begin{equation}
\phi_N  = 2\phi_{N-1} - \phi_{N-2} + \phi_{N-3}\text{,\qquad}N\ge 4\, .
\label{eq:phi_n_extralink_f1}
\end{equation}
This relation can be explained as follows: A sequence of length $N$ is
constructed by inserting a 1 or a 0 after the first 1 in a sequence of
length $N-1$ (giving $\phi_N = 2\phi_{N-1}$). If there was another 1
after the first 1, insertion of a 0 is forbidden. The number of such
forbidden sequences is $\phi_{N-2}$, since they are obtained by
inserting a 1 after the first 1 in a sequence of length $\phi_{N-2}$.
We therefore have to subtract $\phi_{N-2}$. In order to construct
sequences where the first 1 is followed by 001, we insert these 3 bits
after the first 1 in a sequence of length $N-3$.  This means that we
have to add $\phi_{N-3}$. Sequences that contain all 0 or one 1 at the
end are constructed from the all 0 sequence of length $N-1$ by
inserting a 0 or a 1.  Taking into account the boundary condition, the
starting values are $\phi_1=0,\phi_2=3,\phi_3=5$.  The approximate
solution of Eq.~(\ref{eq:phi_n_extralink_f1}), valid with the
precision of the numerical evaluation for $N\lesssim 20$, is
$\phi_N=0.75488\exp(0.5624N)$. The total number of cycles for $L=1$
can now be estimated as $\phi_N/(2N-1)$. For larger $L$, the periodic
sequence of node values at distance $L$ passes the node $\Sigma$ $L$
times, and the boundary conditions are more involved. We found numerically that
for large $L$ the factor in the exponent is smaller.
Figure \ref{fig:f1_numattr} shows the
number of cycles, averaged over $L$, obtained using computer
simulations. The asymptotic increase is not yet visible for these
small values of $N$. 

\subsection{Case 4: Boolean function $f_{11}$ at $\Sigma$}
\label{subsec:f11}

The last canalyzing function that we want to consider, produces $1$ if
the first input is $1$ and inverts the second input otherwise. This
means that the update rule gives $0$ if and only if the first input is
$0$ and the second one is $1$. The system has a fixed point with all
states being 1. Let us consider the fate of a node value $1$ on a
cycle as we iterate the network. This 1 moves from site $x$ to site
$x-1$ during one time step. When it reaches the node $G_1$ it produces
a $1$ at $\Sigma$.  Thus, a $1$ comes back to its original place after
$N-L$ steps. Similarly, a 0 comes back to its original place after
$N-L$ steps, if there was a 1 at this place $L$ time steps before. We
now show that the period of a cycle is indeed $N-L$ (or a divisor
thereof) by demonstrating that at each site there must be a 1 $L$ time
steps after a 0. Consider site $\Sigma$, and assume that its state is
0. This 0 can only have been produced if there is a 0 at site $L$. $L$
time steps later, there must consequently be a 1 at site $\Sigma$.

In order to estimate the number of cycles, let us consider the
sequence of states at $G_1$ every $L$ time steps for $N$ such time
intervals. For $L=1$ the number of states on cycles is equal to the
number of these sequences with an appropriate boundary condition.
Such sequences have no two 0s next
to each other and their number satisfies the recursion relation
$\phi_N = \phi_{N-1} + \phi_{N-2}$, since a sequence of length $N$ can
be generated either by adding a 1 after the first 0 in the sequence of
length $N-1$ or by adding a 10 after the first 0 in a sequence of
length $N-2$. The recursion relation can be shown to hold for
$L\ll N$, only a prefactor of the solution changes. Note that the
recursion relation is identical to the one in the case of the Boolean
function $f_4$. The total number of cycles diverges therefore as
$e^{0.48121 N}/N$, just as before.

\begin{figure}
\begin{center}
\vskip 1cm
\includegraphics*[width=0.85\columnwidth]{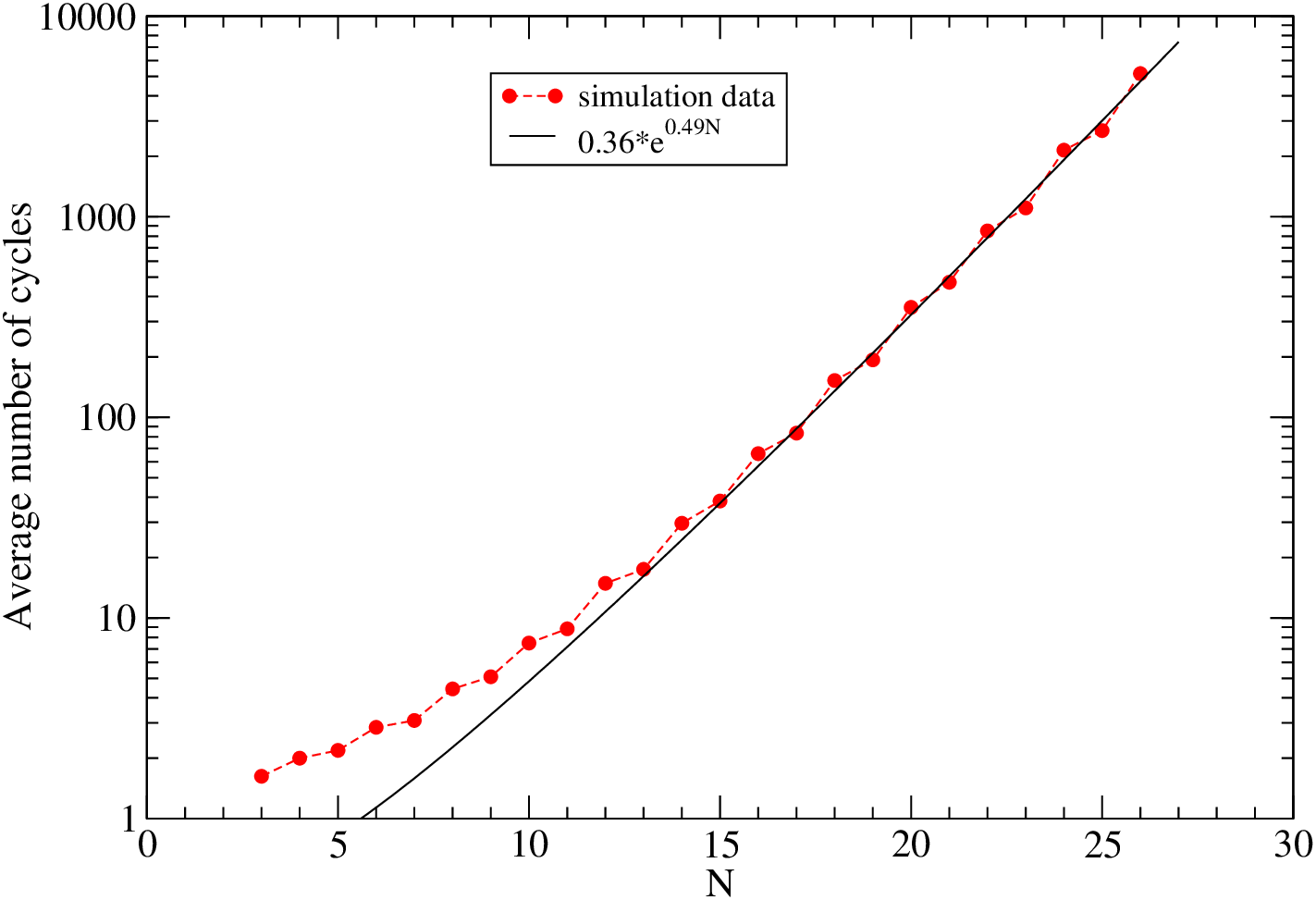}
\end{center}
\caption{Mean number of cycles per network for a canalyzing function
 at $\Sigma$. }
\label{fig:avrg_canal_num}
\end{figure}

The results for all four canalyzing functions indicate that the mean
number of cycles per network, averaged over all canalyzing functions
and values of $L$, should increase at least as fast as $e^{0.5624
N}/N^2$, since a fraction of the order $1/N$ of all networks of size
$N$ have of the order of $e^{0.5624 N}/N$ cycles. However, this
behavior is not yet visible for the $N$ values used in our computer
simulations shown in figure \ref{fig:avrg_canal_num}.

\subsection{Case 5: Boolean function $f_9$ at $\Sigma$}
\label{subsec:f9}

\begin{figure}
\vskip 1.5cm
\begin{center} \setlength{\unitlength}{1pt} 
\begin{picture}(0,0)
\psset{unit=1pt,fillcolor=black}
\pscircle(0,0){40}

\pscircle(-20,34.64){3}
\psline{->}(-14.7,36.7)(-17.35,36.04)
\pscircle(0,40){3}

\pscircle(20,-34.64){3}
\psline{->}(0,-37)(-19.22,31.74)
\pscircle(0,-40){3}

\rput[rt](-45,5){$M^\prime$}
\rput[lt](45,5){$L^\prime$}
\rput[rb](-23,40){$G_2:\Sigma^{-1}$}
\rput[lt](23,-40){$G_1$}
\rput[lt](-3,-45){$G_1^\prime$}

\end{picture}
\end{center}
\vskip 1.5cm
\caption{The network corresponding to the time reversed network in Fig. \ref{fig:extralinkloop} for a reversible Boolean function at $\Sigma$}
\label{fig:revers1}
\end{figure}
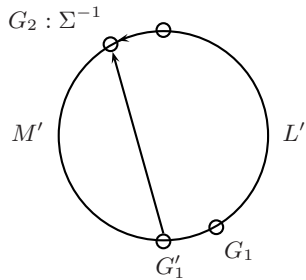

If the Boolean function at $\Sigma$ is reversible, the dynamics on the
system is reversible. All states are on cycles. 
Since  a network with $L\leq M+2$ maps on a network with
$L> M+2$ under time reversal, it is 
sufficient to consider the case $L\leq M+2$, or equivalently
\begin{equation}
\label{eq:Lrestr}
1\leq L\leq [N/2]
\end{equation}
Figure \ref{fig:revers1} shows the time reversed network, with
\begin{eqnarray}
\label{eq:Lprime}
L^\prime &=& M + 2 = N - L\nonumber\\
M^\prime &=& L - 2
\end{eqnarray}
Negative values correspond to self links.

\begin{figure}
\begin{center}
\vskip 1cm
\includegraphics*[width=0.85\columnwidth]{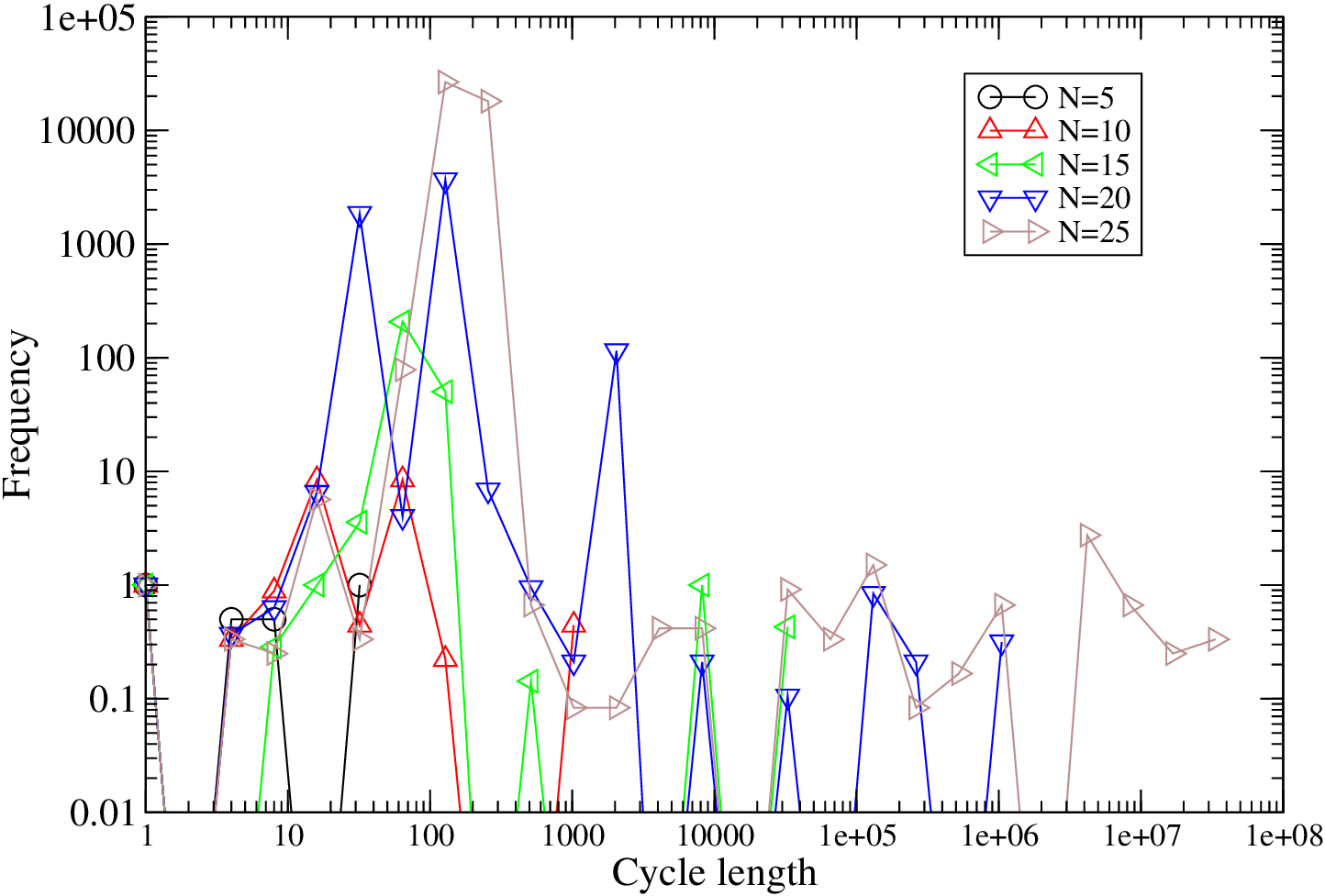}%
\end{center}
\caption{Number of cycles within intervals $[2^n,2^{n+1}]$ for a
reversible Boolean function at $\Sigma$, for selected values of $N$,
averaged over the possible values of $L$.}
\label{fig:p_revers_attr_len_hist}
\end{figure}

\begin{figure}
\begin{center}
\vskip 1cm
\includegraphics*[width=0.85\columnwidth]{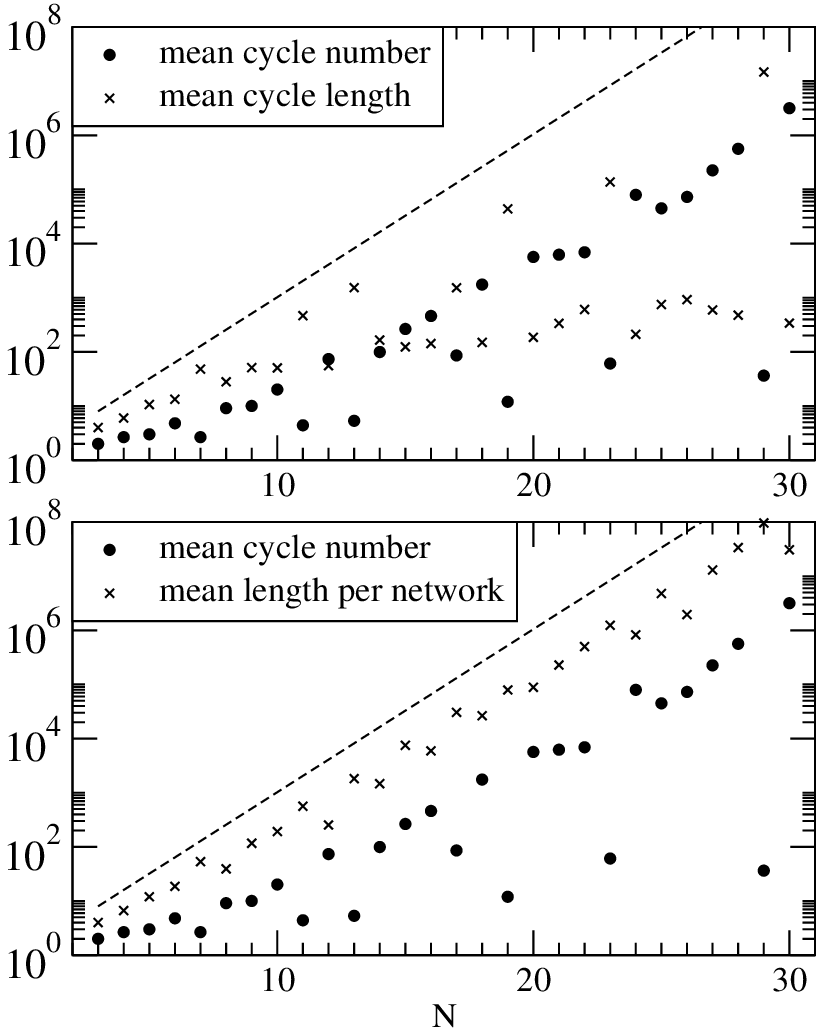}%
\end{center}
\caption{Mean cycle number and length as function of $N$. Top: The cycle length is averaged over all cycles of all systems. Bottom: The mean cycle length is first evaluated for each system separately, before averaging over all systems. The dashed line is the function $2^N$.}
\label{fig:p_revers_attr_len_num}
\end{figure}

If $g$ is the greatest common divisor of $N$ and $L$, the set of all
nodes splits into $g$ independent subsystems with $N/g$ nodes, just as
for the canalyzing functions. In contrast to the canalyzing functions,
each state is now part of a cycle.  The most striking finding is that
there occur now cycles of a length of the order of $2^N$. Figure
\ref{fig:p_revers_attr_len_hist} shows the result of computer
simulations for different values of $N$. One can see that for each of
these $N$ values, there exist cycles of a length close to $2^N$.
Figure \ref{fig:p_revers_attr_len_num} shows the mean number and
length of cycles as a function of $N$. The mean cycle number $$\bar
C_N = \frac{1}{N-1} \sum_{L=1}^{N-1} C_{N,L}$$ shows an exponential increase
for $N$ values that are not prime numbers. The mean cycle length $\bar
P_{N}$ can be defined in different ways:\\
(a) As the mean over all cycle lengths of all systems,
$$\bar P_{N}^{(1)} = \frac{\sum_LC_{N,L}\bar P_{N,L} }{\sum_LC_{N,L}}\, .$$
With this definition, we obtain 
\begin{equation}\bar P_{N}^{(1)} \bar C_N = 2^N\, .\label{eq:product}\end{equation}
This dependence can clearly be seen in the top part of Figure \ref{fig:p_revers_attr_len_num}, where the mean cycle length is largest when $N$ is a prime number and when the cycle number is smallest.\\
(b) As the mean cycle length of a system, averaged over $L$,
$$\bar P_{N}^{(2)} = \frac 1 L \bar P_{N,L}\, .$$
This definition is more physical, since each system should be given the same weight. With this definition, the mean cycle length increases exponentially for all $N$, as shown in the bottom part of Figure \ref{fig:p_revers_attr_len_num}.

A third possible definition of the mean cycle length, which assigns to each possible initial state the same weight, leads to even larger values. 

The occurrence of extremely long periods in systems like these has
been known for some time and has been used in a certain
class of random number generators, see \cite{Mar}, the so-called \emph{Additive Lagged
Fibonacci Generators}. In these random number generators, a sequence
of $m$-bit numbers $x_k$ is generated by the rule
$$x_k = x_{k-p}+ x_{k-p+q} \ {\rm mod}\  m\, .$$ Setting $m=1$, $p=N$, $q = L$,
and using the reversible function $f_6$, this rule gives the sequence
of bits generated at node $\Sigma$ in our network.

\subsection{General considerations}
\label{subsec: general}

We conclude this section by deriving some general results for the
numbers and lengths of cycles in our simple networks. First, we find a
lower bound for the number of cycles for a loop with an extra link for
certain values of $N$. We start with 
$$ C_{2N}^{2L} \ge C_{N}^{L}\cdot C_{N}^{L}/2 \, .$$ The system splits
into 2 independent subsystems, and the inequality arises because the
cycles of the subsystems can have several values of the phase
difference, if their periods have a common divisor. Iterating this equation gives 
$$ C_{2^\nu N_0}^{2^\nu L_0} \ge \left(C_{N_0}^{L_0}/\sqrt{2}\right)^{2^\nu}\equiv C_0^{2^\nu N_0} = C_0^{ N}\, .$$
Now, a given value of $L$ occurs with probability $1/N$ in a system of size $N$, and therefore the mean number of cycles in a system of size $N=2^\nu N_0$ satisfies the inequality
\begin{equation}
C_N \ge \frac 1 N (C_0)^N \equiv 2^{A  N}/N\, .
\end{equation}
The number of cycles increases exponentially with $N$.

Next, we note that we find always an average of one fixed point per
network. For a canalyzing Boolean function at $\Sigma$, we  always
find an average number of 1/4 cycles of length 2. The first
finding can be understood in the following way. If we look at the
state space and consider the ensemble of all networks of size $N$ with
all combinations of update functions, the successor of a state will be with equal probability
every possible state, including itself. The probability that a state is a
fixed point is therefore $1/2^N$. Summing over all states gives an
average of one fixed point.

Now let us consider cycles of length $2$. First of all, there are
no such cycles with reversible update rules. As a matter of fact,
depending on $L$, for the inputs of $\Sigma$ on a cycle of length
$2$ there exist only two possibilities: they alternately take on the
values $(0,1)$ and $(1, 0)$ or they alternate between $(0, 0)$ and $(1,
1)$. In both cases the output of the reversible function would be
constant, thus leaving no space for a cycle of length $2$.

We turn to the canalyzing functions. Simulation data show that on
average every fourth network has a cycle of length $2$. We want to
give two different proofs for this.  Consider a state $g^{(i)}$. As
with the fixed points, the statistical probability that $g^{(i)}$ is
followed by $g^{(j)}$ under the dynamics, is $1/2^N$.  We denote the
corresponding set of networks that make this transition by ${\mathcal
N}_{ij}$. The question now is, what is the probability that the state
$g^{(j)}$ returns to $g^{(i)}$ in the next step. For the networks in
${\mathcal N}_{ij}$ to perform the transition $i\mapsto j$ for fixed
$i$ and $j$ the update rule at $\Sigma$ is fixed for one of $4$ input
states, thus ruling out 4 out of the 8 canalyzing Boolean
functions. Thus the probability, that ${\mathcal N}_{ij}$ leads
$g^{(j)}$ to $g^{(i)}$ at the next time step is $4/(8\cdot 2^N)$.
Altogether we get the following result for the probability $p_2$ of a
cycle of length $2$:
\begin{equation}
\label{eq:p2}
p_2 = \frac{1}{2}\sum_{i,j}\frac{1}{2^N}\frac{1}{2\cdot 2^N} = 1/4
\end{equation}

We can also see this directly, by constructing explicitely the cycles
of length $2$. These cycles are sequences of alternating $0$ and $1$s,
which have two 0s (or two 1s) together at $\Sigma$ for odd
$N$s. Without loss of generality, we use only truth functions as
update rules at nodes with one input. $\Sigma$ has either inputs
alternating between $0,1$ and $1,0$ for odd $L$, or inputs alternating
between $0, 0$ and $1, 1$ for even $L$, and the output must be
alternating 0s and 1s. In each mentioned case, for any fixed $N$ and
$L$, two of eight canalyzing functions are suitable. For example, for
an odd $N$ and odd $L$ the output for the input state $01$ (the right
value is the first input), has to be $1$; it has to be $0$ for
$10$. Thus for all $L$s and for all possible update rules the fraction
of networks with a length $2$ cycle is $2/8=1/4$.

\section{Conclusions}
\label{sec:conclusions}

In this paper, we have investigated mainly analytically the effect of adding one
additional link to networks consisting of one or two simple loops. There was a
big difference in the typical numbers and lengths of cycles between
networks with a canalyzing Boolean function and networks with a
reversible Boolean function. For two loops with a cross-link, a
reversible coupling function between the two loops leads to results
very similar to those for two independent loops. However, a canalyzing
function reduces the typical values of cycle length and number to
those of a single loop. One gets an increased number of cycles for $N_1=N_2$.
For canalyzing functions one finds several dominant cycle lengths.

For loops with an additional link, one of the canalyzing functions can
freeze the entire network, while other canalyzing functions produce
cycles of a period up to $2N$. The number of cycles increases
exponentially with $N$, but not as fast as for simple loops. The most
interesting finding was that a reversible function generates mean
cycle lengths that increase exponentially with the network size. 

We thus have shown that even very simple networks consisting of
relevant nodes with reversible couplings have a mean cycle length and
a mean cycle number that increase faster than any power law in network
size. On the other hand, canalyzing couplings tend to reduce the cycle
length and number compared to the case where the additional link is
absent. It will be interesting to see how these two contrary effects
of canalyzing and reversible couplings work together in more
complicated relevant components of larger networks. 

Our calculations give some indications for why it is so difficult to
measure correct values for cycle numbers and lengths in computer
simulations of critical Kauffman networks. Even for the simple
components considered in this paper, there are cycles that can only be
reached from a small fraction of initial conditions. For instance, in
the case of two loops with a cross-link, many cycles have a frozen
first loop. However, these cycles are only reached from initial
conditions with a frozen first loop, which are a fraction of the order
$2^{-N_1}$ of all initial conditions. Furthermore, for combinations of
$N_1$ and $N_2 $, or of $N$ and $L$, which have many common divisors,
there exist particularly large numbers of cycles. By sampling only a
small number of initial conditions, it will never be possible to find
all these cycles. For these reasons, we have always performed a
complete search of state space in the simulations reported in this
paper. 

The findings of this paper teach us a third lesson: Even with a
thorough exploration of state space, it can be difficult to see the
true asymptotic behavior of mean cycle numbers or sizes, as
demonstrated in the case of a loop with an additional link and with a
canalyzing coupling. Different contributions for different coupling
functions and for different values of $L$ can increase in a different
way with $N$. The contribution that increases fastest will only
dominate if $N$ becomes very large. Only then will the true asymptotic
behavior become visible. 

One of the main conclusions of these findings is that a purely
numerical investigation of Kauffman networks will never produce
reliable results. It is essential to develop analytical approaches
that help to understand the important features of these systems. Up to
now, there exist few analytical studies, and many more will be needed
before Kauffman networks will be fully understood.

\bibliography{litverz}


\end{document}